\documentclass[twocolumn,english,prl,showpacs,aps]{revtex4}
\usepackage[T1]{fontenc}
\usepackage[utf8]{inputenc}
\usepackage{amssymb}
\usepackage{esint}
\usepackage{times}
\usepackage{graphicx}
\usepackage{color}
\usepackage[normalem]{ulem} % Marc - Pour pouvoir barrer du texte.

\newcommand{\krec}{\ensuremath {k_{\rm rec}}}

\begin{document}

\title{The 2nd order coherence of superradiance from a Bose--Einstein condensate}

\author{R.~Lopes}
\author{A.~Imanaliev}
\author{M.~Bonneau}
\altaffiliation{Current address: Vienna Center for Quantum Science and Technology, Atominstitut, TU Wien, Stadionallee 2, 1020 Vienna, Austria.}
\author{J.~Ruaudel}
\altaffiliation{Current address: Laboratoire Kastler Brossel, Univ. Pierre et Marie Curie - École normale supérieure - CNRS, 4 place Jussieu, 75005 Paris, France.}
\author{M.~Cheneau}
\author{D.~Boiron}
\author{C.~I.~Westbrook}
\email{christoph.westbrook@institutoptique.fr}
\affiliation{Laboratoire Charles Fabry, Institut d'Optique - CNRS - Univ. Paris-Sud, 2 avenue Augustin Fresnel, 91127 Palaiseau, France}

\date{\today}

\begin{abstract}
{We have measured the 2-particle correlation function of atoms from a Bose--Einstein condensate participating in a superradiance process, which directly reflects the 2nd order coherence of the emitted light.
We compare this correlation function with that of atoms undergoing stimulated emission.
Whereas the stimulated process produces correlations resembling those of a coherent state, we find that superradiance, even in the presence of strong gain, shows a correlation function close to that of a thermal state, just as for ordinary spontaneous emission.
}
\end{abstract}

\maketitle

%%%%%%%%%
%    Introduction 
%%%%%%%%%

Ever since the publication of Dicke's 1954 paper \cite{Dicke:1954}, the problem of the collective emission of radiation has occupied many researchers in the fields of light scattering, lasers and quantum optics.
Collective emission is characterized by a rate of emission which is strongly modified compared to that of the individual atoms \cite{Gross:1982}. It occurs in many different contexts: hot gases, cold gases, solids and even planetary and astrophysical environments \cite{Letokhov:2008}.
The case of an enhanced rate of emission, originally dubbed superradiance, is closely connected to stimulated emission and gain, and as such resembles laser emission \cite{Siegman:1986}.
Lasers are typically characterized by high phase coherence but also by a stable intensity, corresponding to a Poissonian noise, or a flat 2nd order correlation function \cite{Loudon:2000}.
Here we present measurements showing that the coherence properties of superradiance, when it occurs in an ultracold gas and despite strong amplified emission, are much closer to those of a thermal state, with super-Poissonian intensity noise.

Research has shown that the details of collective emission depend on many parameters such as pumping configuration, dephasing and relaxation processes, sample geometry, the presence of a cavity, etc. and, as a result, a complex nomenclature has evolved including the terms superradiance, superfluorescence, amplified spontaneous emission, mirrorless lasing, and random lasing  \cite{Rehler:1971, Allen:1973, Bonifacio:1975, MacGillivray:1976, Gross:1982, Siegman:1986}, the distinctions among which we will not attempt to summarize here.
The problem has recently seen renewed interest in the field of cold atoms \cite{Inouye:99a,Moore:1999, Mustecaplioglu:2000, Piovella:2003, Pu:2003, Yoshikawa:2004, Bar-Gill:2007, Wang:2007, Paradis:2008, Hilliard:2008, Meiser:2010, Deng:2010, Vogt:2011, Bohnet:2012, Greenberg:2012, Baudouin:2013}.
This is partly because cold atoms provide a reproducible, easily characterized ensemble in which Doppler broadening effects are small and relaxation is generally limited to spontaneous emission.
Most cold atom experiments differ in an important way from the archetypal situation first envisioned by Dicke: instead of creating an ensemble of excited atoms at a well defined time and then allowing this ensemble to evolve freely, the sample is typically pumped during a period long compared to the relaxation time and emission lasts essentially only as long as the pumping.
The authors of reference \cite{Inouye:99a} however, have argued that there is a close analogy to the Dicke problem, and we will follow them in designating this process as superradiance.

In the literature on superradiance there has been relatively little discussion about the coherence and correlation properties of the light. 
The theoretical treatments we are aware of show that the coherence of collective emission can be quite complicated, but does not resemble that of a laser \cite{Haake:1972, Gross:1982, Moore:1999a, Piovella:2003, Temnov:2009, Meiser:2010}. These results, however, were obtained for simple models that do not include all parameters relevant to laboratory experiments. Experimentally, a study performed on Rydberg atoms coupled to a millimeter-wave cavity \cite{Raimond:1982} showed a thermal mode occupation, and an experiment in a cold atomic vapor in free space \cite{Greenberg:2012} observed a non-flat 2nd order correlation function. 
In the present work, we show that even if the initial atomic state is a Bose--Einstein condensate (BEC), the 2nd order correlation function looks thermal rather than coherent.

Such behavior, which may seem counter-intuitive, can be understood by describing superradiance as a four wave mixing process between two matter waves and two electromagnetic waves.
The initial state consists of a condensate, a coherent optical pump beam, and empty modes for the scattered atoms and the scattered photons.
If we make the approximation that the condensate and the pump beam are not depleted and can be treated as classical fields, the matter-radiation interaction Hamiltonian is given by:
\begin{equation}
\hat{H} = \sum_i \left[ \chi_i \, \hat{a}^\dag_{\text{at},i} \, \hat{a}^\dag_{\text{ph},i} + \chi_i^{\ast} \, \hat{a}_{\text{at},i} \, \hat{a}_{\text{ph},i} \right] \; ,
\label{hamiltonian}
\end{equation}
where $\hat{a}^\dag_{\text{at},i}$ ($\hat{a}_{\text{at},i}$) and $\hat{a}^\dag_{\text{ph},i}$ ($\hat{a}_{\text{ph},i}$) denote atom and photon creation (annihilation) operators for a specific pair of momenta $i$ fixed by energy and momentum conservation and $\chi_i$ is a coupling constant.
Textbooks \cite{Gerry:2005} show that, starting from an input vacuum state, this Hamiltonian leads to a product of two-mode squeezed states. When one traces over one of the two modes, $\alpha = \{\text{at},i\}$ or $\{\text{ph},i\}$, the remaining mode $
\beta$ has a thermal occupation with a normalized 2-particle or 2nd order correlator
\begin{equation}
\frac{\langle \hat{a}^\dag_{\beta} \, \hat{a}^\dag_{\beta} \, \hat{a}_{\beta} \,  \hat{a}_{\beta} \rangle}{\langle \hat{a}^\dag_{\beta} \, \hat{a}_{\beta} \rangle^{2}} = 2 \; ,
\end{equation}
whereas it is unity for a laser.
The problem has also been treated for four wave mixing of matter waves \cite{Molmer:2008}.
We emphasize that, when starting from initially empty modes, the occupation remains thermal {\it regardless} of the gain.

In the experiment, we start from initially nearly motionless atoms of a BEC and observe their recoil upon photon emission.
To the extent that each recoil corresponds to the emission of a single photon, we can obtain essentially the same information about the radiation from such measurements as by observing it directly. 
In doing this, we are following the approach pioneered in experiments such as \cite{Raimond:1982, Inouye:99a} and followed by many others, which uses highly developed atom detection and imaging techniques to glean most of the experimental information about the process. 
We are able to make time-integrated measurements of the emission, resolved in transverse and longitudinal momentum as well as in polarization, and reconstruct the 2-particle correlation function of the recoiling atoms, or equivalently the 2nd order correlation function of the scattered light.
We will show that in the configuration of our experiment, the 2nd order correlation is close to that of a thermal sample, and very different from the correlation properties of the initial, condensed atomic state.

%%%%%%%%%%%%
% Experiment
%%%%%%%%%%%%

\begin{figure}[t]
\begin{center}
\includegraphics[width=\columnwidth]{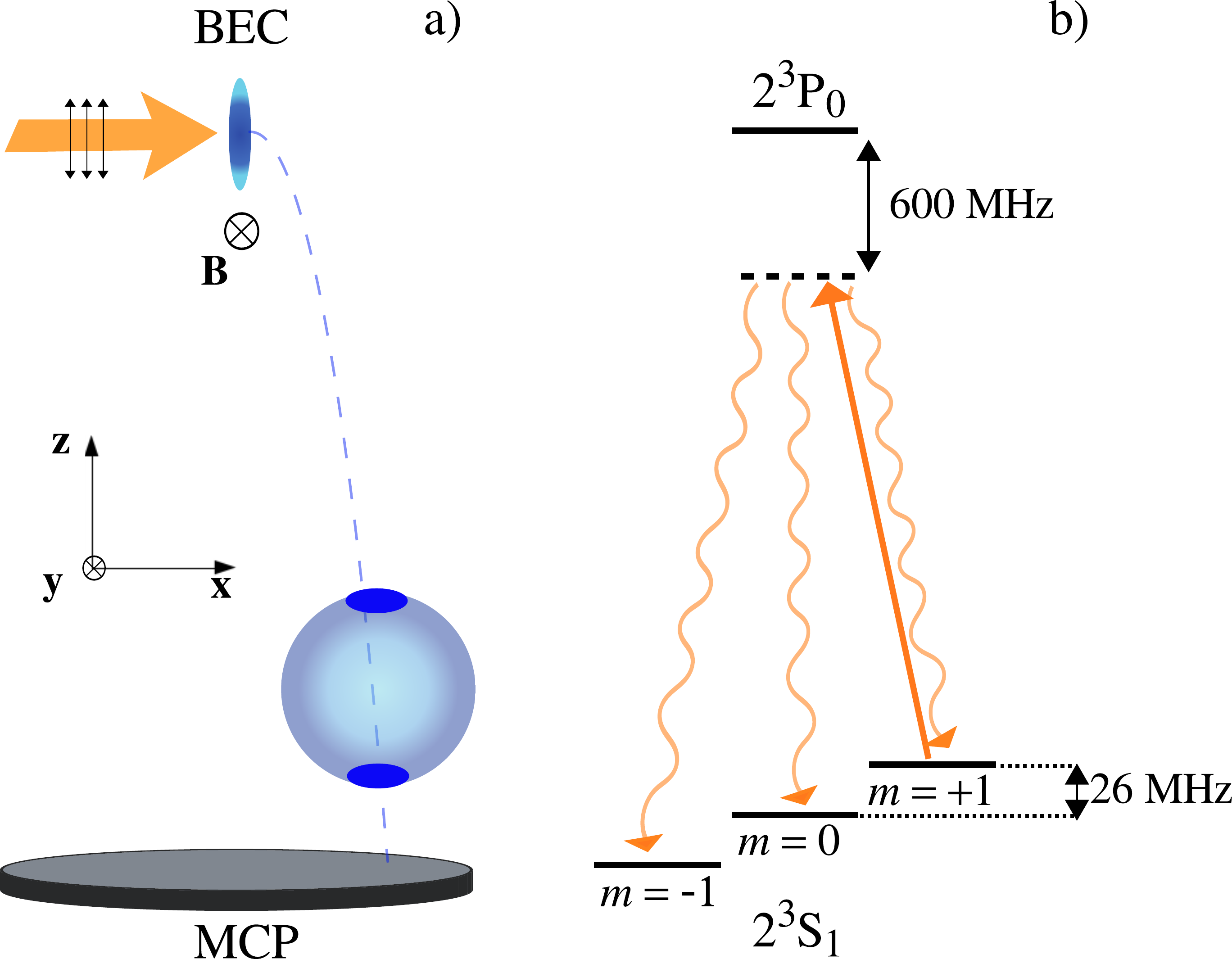}
\caption{(color online) (a) Sketch of the experiment. 
A 9-G magnetic field $\mathbf{B}$ applied along the $y$ axis defines the quantization axis. The excitation beam propagates with an angle of $10^\circ$ (not shown) relative to the $x$ axis and its polarization is linear, with the same angle relative to the $y$ axis. 
After emission, the atoms fall 46\,cm to a position-sensitive micro-channel plate (MCP). 
The atom cloud  forms a sphere with enhanced occupation of the endfire modes. 
(b)  Atomic level scheme. 
The atoms, initially in the $2^3S_1,\,m=+1$ state, are excited to the $2^3P_0$ state. From there, they can decay with equal branching ratios to the 3 sub-levels of the ground state. We detect only the atoms which scatter into the $m=0$ state.}
\label{experiment}
\end{center}
\end{figure}

We use helium in the $2^3S_1,\,m=1$ state confined in a crossed dipole trap (see Fig.~\ref{experiment}a) with frequencies of 1300~Hz in the $x$ and $y$ directions and  130\,Hz in the (vertical) $z$ direction.
The dipole trap wavelength is $1.5\,\mu$m. 
The atom number is approximately 50\,000 and the temperature of the remaining thermal cloud 140\,nK. 
A 9~G magnetic field along the $y$ axis defines a quantization axis. 
After producing the condensate, we irradiate it with a laser pulse of 2.4\,W/cm$^2$  tuned  600\,MHz to the red of the $2^{3}S_{1}\rightarrow2^{3}P_{0}$ transition at $\lambda= 1083$\,nm and with natural linewidth 1.6\,MHz. 
The excitation beam propagates with an angle of $10^{\circ}$ relative
to the $x$ axis and its polarization is linear, with the same angle relative to the $y$ axis (see Fig.~\ref{experiment}a).
The pulse length is $5\,\mu$s and it is applied with a delay $\tau$ after switching off the trap.
The expansion of the cloud during this delay is a convenient way to vary both the optical density and the anisotropy of the sample at constant atom number.
The absorption dipole matrix element is of the $\sigma^-$ form and thus one half of the laser intensity is coupled to the atomic transition corresponding to a Rabi frequency of 56\,MHz.
The excited atoms decay with equal branching ratios to the 3 ground states. 
During the pulse, less than 10\,\% of the atoms are pumped into each of these states. 
Because of the polarization selection rules, the atoms which are pumped into the $m=0$ state cannot reabsorb light from the excitation laser.
By focusing on these atoms, we study the regime of ``Raman superradiance'' \cite{Schneble:2004, Yoshikawa:2004}, by which we mean that an absorption and emission cycle is accompanied by a change of the internal state of the atom.
When the trap is switched off, the atoms fall toward a micro-channel plate detector which detects individual atoms with 3 dimensional imaging capability and a 10 to $20\,\%$ quantum efficiency \cite{Jaskula:2010}.
A magnetic field gradient is applied to sweep away all atoms except those scattered into the $m=0$ magnetic sublevel.
The average time of flight to the detector is 310\,ms and is long enough that the atoms' positions at the detector reflect the atomic momenta after interaction with the excitation laser.
Conservation of momentum then requires that these atoms lie on a sphere with a radius equal to the recoil momentum $\krec = 2\pi/\lambda$.
Any additional scattering of light, whether from imperfect polarization of the excitation laser, or from multiple scattering by the atoms, will result in the atoms lying outside the sphere. We see no significant signal from such events, but in order to completely eliminate the possibility for multiple scattering we restrict our analysis of the data to the spherical shell with inner radius $0.8\,\krec$ and outer radius 1.2\,\krec.

We excite atoms in an elongated BEC in such a way that an allowed emission dipole can radiate along the long axis. In an anisotropic source, collective emission builds up more efficiently in the directions of highest optical thickness. Superradiance is therefore expected to occur along the long axis of the BEC, in so called ``endfire'' modes \cite{Dicke:1964, Inouye:99a}.
An important parameter then is the Fresnel number of the sample \cite{Gross:1982}, $F={2 R_\perp^2 / \lambda R_z}$, where $R_{\perp}$ and $R_z$ are the horizontal and vertical Thomas--Fermi radii of the condensate. The Fresnel number distinguishes between the diffraction limited ($F<1$) and multimode superradiance regimes ($F>1$). In our case, $R_{\perp}\approx 5\,\mu$m and $R_z\approx 50\,\mu$m, yielding a Fresnel number of about unity.

\begin{figure}[t]
\begin{center}
\includegraphics[width=\columnwidth]{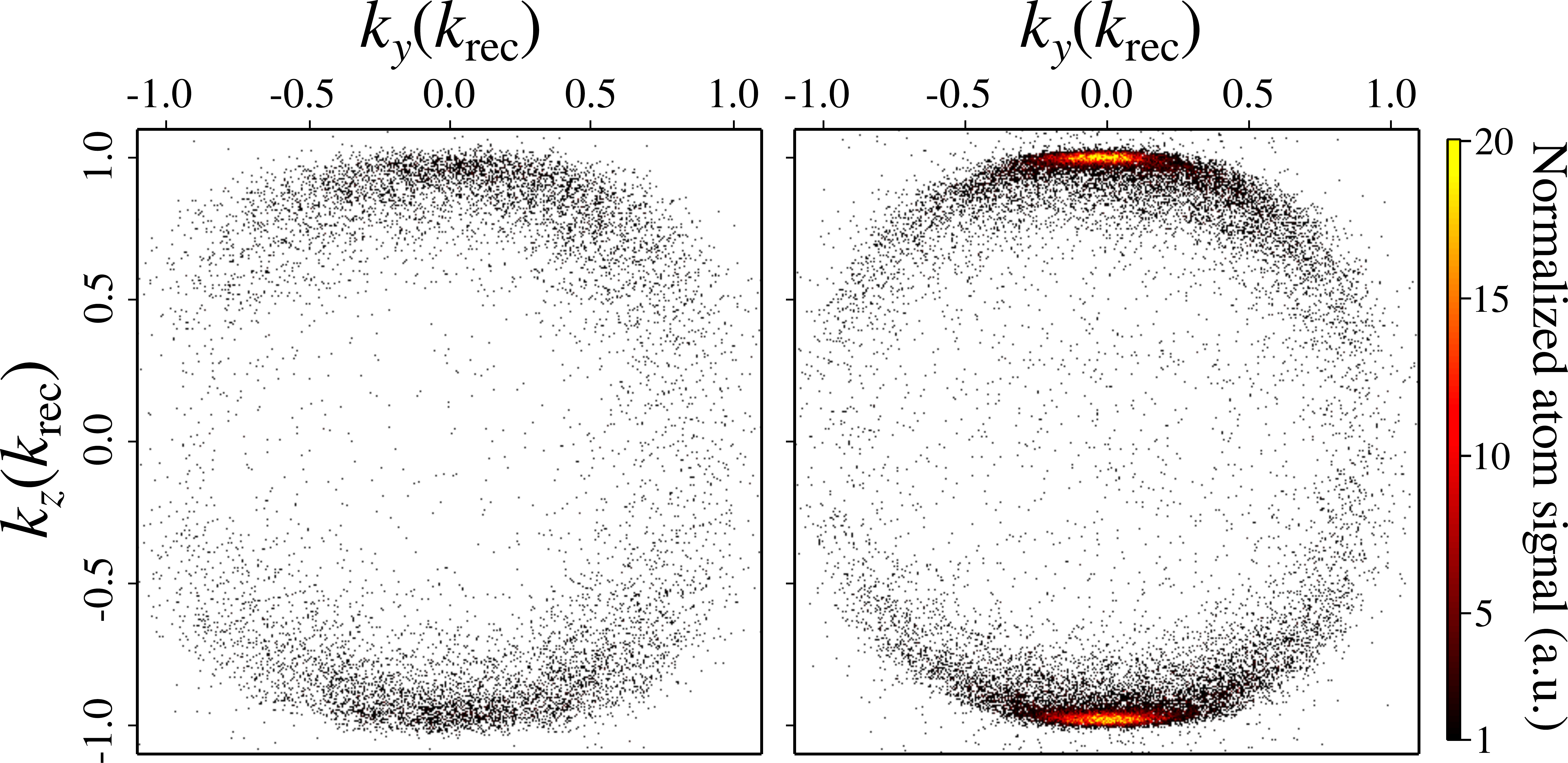}
\caption{(color online) Momentum distribution of scattered atoms in the $yz$ plane (containing the emission dipole).
Both figures show the distribution in the $yz$ plane, integrated between $k_x=\pm 0.1\,\krec$ and summed over 100 shots. See the supplemental information for a cut in the $xz$ plane \cite{supp}.
Left: Excitation laser applied 500\,$\mu$s after the trap switchoff. 
Only the radiation pattern for a $y$-polarized dipole is visible.
Right: Excitation laser applied immediately after the trap switchoff. 
Strong superradiance is visible in the vertical, endfire modes. 
}
\label{3D}
\end{center}
\end{figure}

Typical cuts through the atomic momentum distribution in the $yz$ plane are shown in Fig.~\ref{3D}, for $\tau = 500\,\mu$s (left panel) and $\tau \approx 0$ (right panel). 
In both cases, the spherical shell with radius $1\,\krec$ appears clearly. For the short delay, when the atomic sample remains dense and anisotropic, we observe strong scattering in the endfire modes at the top and bottom poles of the sphere.
In addition to this change in the profile of the distribution, we measure an increase of the \emph{total number} of atoms on the sphere by a factor $\sim$ 5 from $\tau = 500\,\mu$s to $\tau \approx 0$. Because each atom has scattered a single photon, this increase directly reflects an increase of the rate of emission in the sample and therefore demonstrates the collective nature of the scattering process.
At long delays, the condensate has expanded sufficiently that the optical thickness and anisotropy have fallen dramatically, suppressing the collective scattering.
By looking at the number of scattered atoms in the $x$ direction (perpendicular to the plane of Fig.~\ref{3D}),
we have verified that, away from the endfire modes, the rate of emission varies by less than 10\% for different delays \cite{supp}.

\begin{figure}[t]
\begin{center}
\includegraphics[width=\columnwidth]{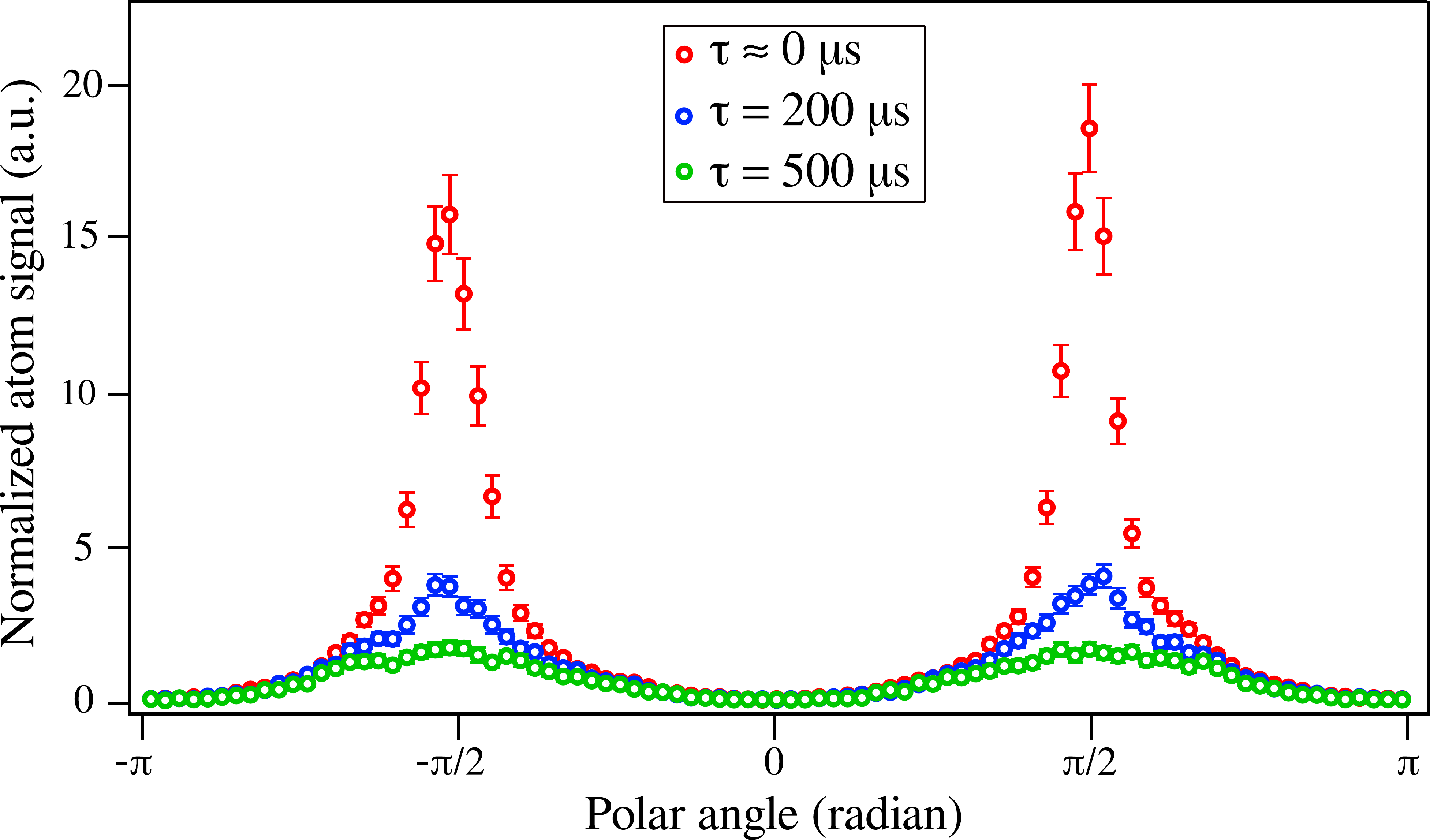}
\caption{(color online) Angular distribution of scattered atoms in the $yz$ plane (containing the emission dipole) for different values of the delay $\tau$ before the excitation pulse.
The data for $\tau =$ 0 and 500\,$\mu$s are the same as those shown in Fig.~\ref{3D}.
The images were integrated along the $x$-axis between $\pm 0.1\,\krec$ and only atoms lying inside a shell with inner radius 0.8 $\krec$ and outer radius 1.2\,$\krec$ were taken into account. 
The delays $\tau=0$, 200 and 500 $\mu$s correspond to peak densities of $\approx 8,\, 2,\, 0.4\times10^{18}\,{\rm m}^{-3}$ and to aspect ratios of 10, 5 and 2.5, respectively.
The endfire modes are located at $\pm \pi /2$.
The half-width at half-maximum of the highest peak is 0.14 rad.
Error bars are shown and denote the $68\,\%$ confidence interval.
}
\label{AngularPlot}
\end{center}
\end{figure}

To see the distribution in a more quantitative way, we show in Fig.~\ref{AngularPlot} an angular plot of the atom distribution in the $yz$ plane.
Data is shown for three different delays $\tau$ before application of the excitation pulse.
For 500\,$\mu$s delay, the angular distribution follows the well-known ``$\sin^2\theta$'' linear dipole emission pattern with the angles $\theta=0$ and $\pi$ corresponding to the orientation of the dipole along the $y$-axis \cite{supp}.
For 200\,$\mu$s delay, the superradiant peaks are already visible on top of the dipole emission profile.
For the shortest delay, the half width of the superradiant peaks is $0.14\, \krec$, or $0.14$ rad, consistent with the diffraction angle and the aspect ratio of the source. 
In the vertical direction, the superradiant peaks are 10 times narrower than in the horizontal direction \cite{supp}.

\begin{figure}[t]
\begin{center}
\includegraphics[width=\columnwidth]{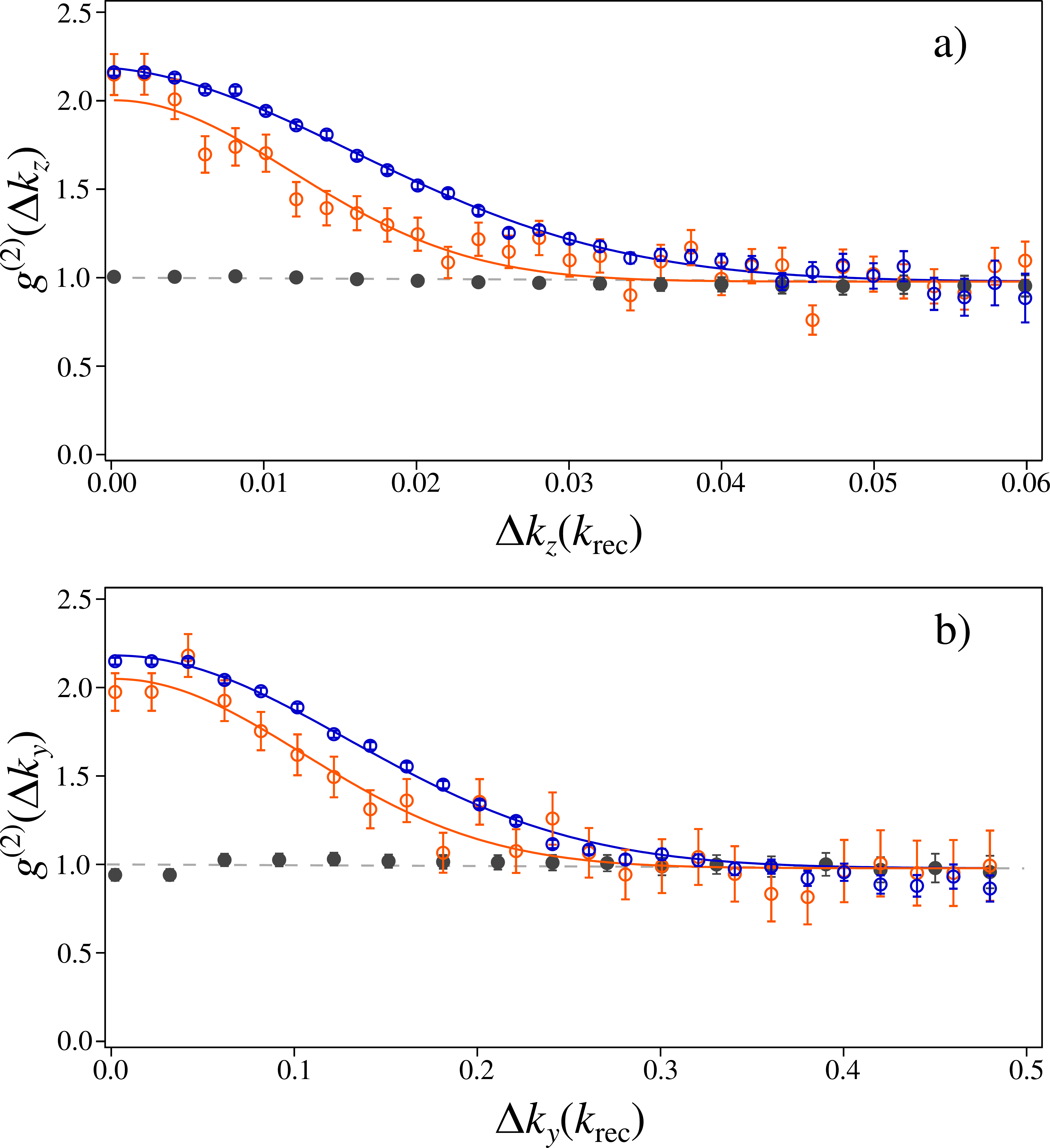}
\caption{(color online) 
Correlation functions along the $z$ (a) and $y$ axis (b) for $\tau \approx 0$.
Blue circles correspond to the superradiant peaks (defined by $|k_z| > 0.95 \krec$).
Orange circles correspond to atoms from the scattering sphere away from the superradiant peaks (defined by $|k_z| < 0.92 \krec$). 
Solid lines are Gaussian fits constrained to approach unity at large separation.
Gray solid circles correspond to a fraction of the initial condensate transferred to the $m=0$ state via a stimulated Raman transfer.
The dashed gray line shows unity. Error bars denote the $68 \%$ confidence interval.
}
\label{Correlations}
\end{center}
\end{figure}

In the strongly superradiant case, we observe large and uncorrelated fluctuations of the heights of the two superradiant peaks on a shot-to-shot basis. These fluctuations directly reflect the fluctuations of the population of the superradiant modes.
We investigate these fluctuations further by measuring the normalized 2-particle correlation function of the scattered atoms, defined as
\begin{equation}
g^{(2)}(\Delta {\bf k})={\langle\,:\!\hat n({\bf k})\hat n({\bf k}+\Delta {\bf k})\!:\,\rangle
\over
\langle \hat n({\bf k})\rangle \;\langle \hat n({\bf k}+\Delta {\bf k}) \rangle} \; .
\end{equation}
Here, $\hat n$ is the atomic density and $: \; :$ denotes normal ordering.  
In practice, this function is obtained from a histogram of pair separations $\Delta {\bf k}$ normalized to the autoconvolution of the average particle momentum distribution \cite{Schellekens:2005, Kheruntsyan:2012}. 
Figure~\ref{Correlations} shows the experimentally measured correlation functions integrated over the momentum along two out of three axes, both for the superradiant peaks and on the scattering sphere away from the peaks \cite{supp}.

We see that in both cases the correlation function at zero separation reaches a value close to 2. This shows clearly that, despite strong amplified emission in the endfire modes, the atoms undergoing a superradiant process have statistics comparable to that of a thermal sample. As underlined in the introduction, these large fluctuations can be simply understood by modeling the superradiant emission as a four wave mixing process; they arise from the fact that the emission is triggered by spontaneous emission. 
For the superradiant peaks, the correlation actually is slightly larger than 2.
Similar behavior has appeared in some models \cite{Meiser:2010, Wasak:2012}, but these models may not be directly applicable to our situation.

Figure \ref{Correlations} also shows that the correlation widths of the superradiant modes are somewhat broader than those of the atoms scattered in other modes.
The effect is a factor of about 1.5 in the vertical direction and about 1.25 in the horizontal direction \cite{supp}. 
The broadening indicates that the effective source size for superradiance is slightly smaller than for spontaneous scattering.
A decreased vertical source size for superradiance is consistent with the observations of Ref.~\cite{Zobay:2006, Sadler:2007} which showed that the superradiant emission is concentrated near the ends of the sample. 
In the horizontal direction, one also expects a slightly reduced source size relative to the atom cloud since the gain is higher in the center, where the density is higher. The fact that the correlation widths are close to the widths of the momentum distribution \cite{supp} indicates that the superradiant peaks are almost single mode as expected for samples with a Fresnel number close to unity \cite{Gross:1982}.

The spontaneous superradiant scattering process should be contrasted with stimulated Raman scattering. 
In terms of the model described by the Hamiltonian (\ref{hamiltonian}), stimulated Raman scattering corresponds to seeding one of the photon modes by a coherent state.
In this case, vacuum fluctuations do not initiate the scattering process, and the resulting mode occupation is not thermal but coherent.
To study stimulated scattering, we applied the excitation beam together with another beam polarized parallel to the magnetic field and detuned by the Zeeman shift (25\,MHz) with respect to the $\sigma$-polarized beam, inducing a stimulated Raman transition.
The laser intensities were adjusted to transfer a similar number of atoms to the $m=0$ state as in the superradiance experiment. 
The normalized correlation functions in this situation, shown in Fig.~\ref{Correlations}, are very nearly flat and equal to unity as we expect for a BEC \cite{Schellekens:2005, Ottl:2005, Hodgman:2011}. The complementary experiment, seeding the {\it atomic} mode with a coherent state has also been observed to produce a coherent amplified matter wave \cite{Inouye:1999, Kozuma:1999}.
As a side remark, we have also observed that the superradiant atom peaks are 2.8 times narrower in the vertical direction than the vertical width of the transferred condensate \cite{supp}. We attribute this to a longitudinal gain narrowing effect \cite{Moore:1999a}.

We also investigated the influence of several other experimental parameters on the 2nd order coherence of the superradiant emission: We have excited the atomic sample with a longer and stronger pulse ($10\,\mu$s, 3.2\,W/cm$^2$), so that the initial condensate was entirely depleted. We have explored the Rayleigh scattering regime, in which the atoms scatter back to their initial internal state.
We also changed the longitudinal confinement frequency of the BEC to 7\,Hz, leading to a much greater aspect ratio. These different configurations led to 2-particle correlation functions which were very similar to the one discussed above.
We believe that similar fluctuations will occur in superradiance from a thermal cloud\, provided that the gain in the medium is large enough. We were unable to confirm this experimentally in our system, precisely because of the vastly reduced optical density.
However, non-coherent intensity fluctuations have been observed already using magneto-optically trapped atoms \cite{Greenberg:2012}.
This seems to confirm our interpretation that the large fluctuations of the superradiant mode occupation is an intrinsic property of superradiant emission, reflecting the seeding by spontaneous emission. The only way to suppress these fluctuations would be to restrict the number of scattering modes to one by means of a cavity and to saturate the gain by completely depleting the atomic cloud. The occupation of the superradiant mode would then simply reflect that of the initial atomic sample.

An interesting extension of the techniques used here is to examine superradiant Rayleigh scattering of a light pulse short enough and strong enough to populate oppositely directed modes \cite{Schneble:2003}. 
It has been predicted \cite{Piovella:2003, Pu:2003, Buchmann:2010} that the modes propagating in opposite direction are entangled, similar to those produced in atomic four wave mixing \cite{RuGway:2011, Bucker:2011, Bonneau:2013}.
A similar measurement technique should be able to reveal them.

\vskip 6pt
\begin{acknowledgments}
We acknowledge fruitful discussions with A. Browaeys, J.-J. Greffet and P. Pillet.
This work was supported by the IFRAF institute, the Triangle de la Physique, the LABEX PALM, the ANR-ProQuP project, the ERC (Grant No. 267 775) Quantatop, J.R. by the DGA, R.L. by the FCT scholarship SFRH/BD/74352/2010.

\end{acknowledgments}

\clearpage
\renewcommand\thefigure{S\arabic{figure}} 
\setcounter{figure}{0}  
\renewcommand\theequation{S\arabic{equation}} 
\setcounter{equation}{0}  
\section{Supplementary material}

\begin{figure}[h]
\begin{center}
\includegraphics[width=\columnwidth]{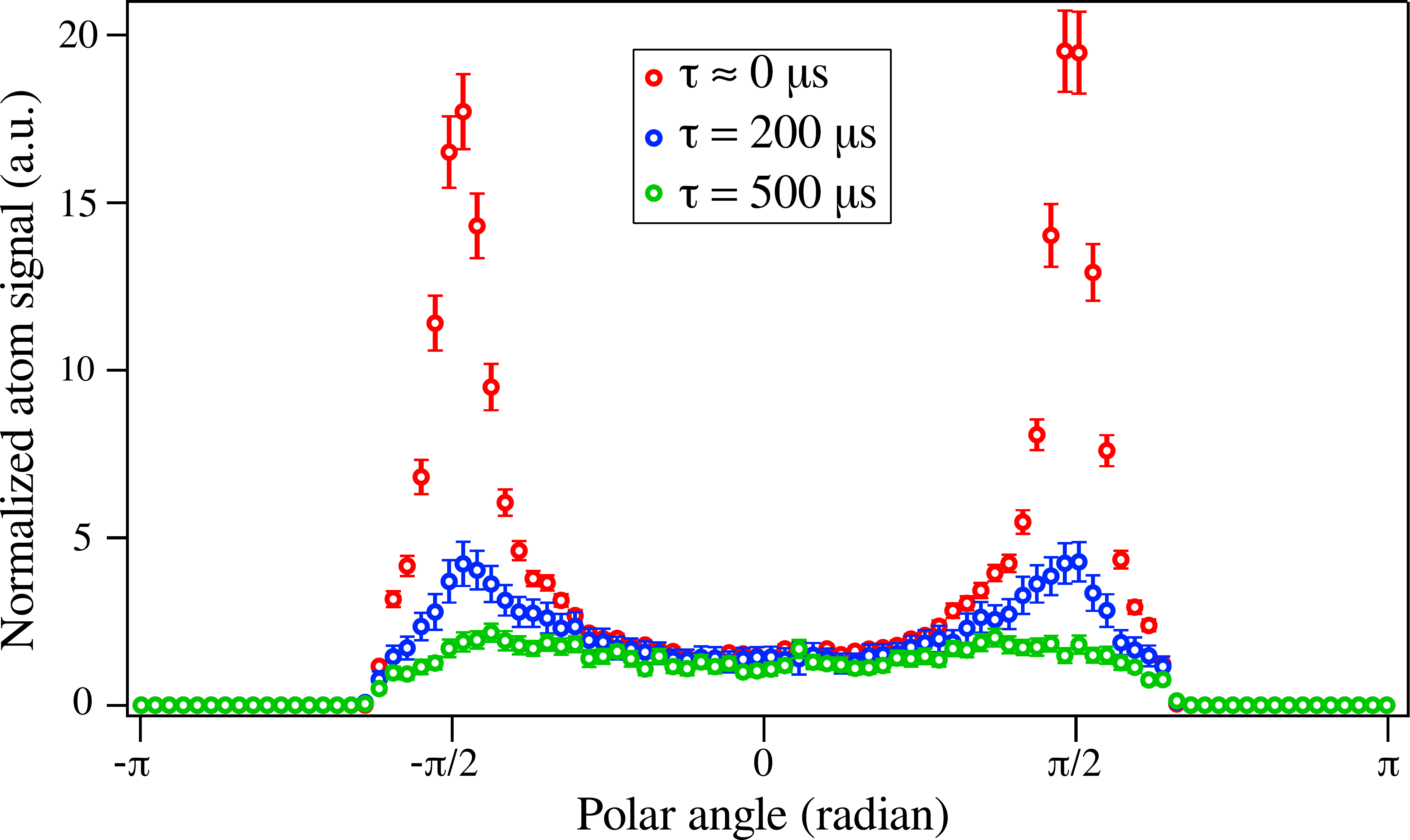}
\caption{(colour online)
Angular distribution of scattered atoms in the plane perpendicular to the emission dipole for different values of the delay $\tau$ before the excitation pulse. The data shown are the same as those discussed in the main text.
}
\label{AngularPlot_X}
\end{center}
\end{figure}

\begin{figure}[h]
\begin{center}
\includegraphics[width=\columnwidth]{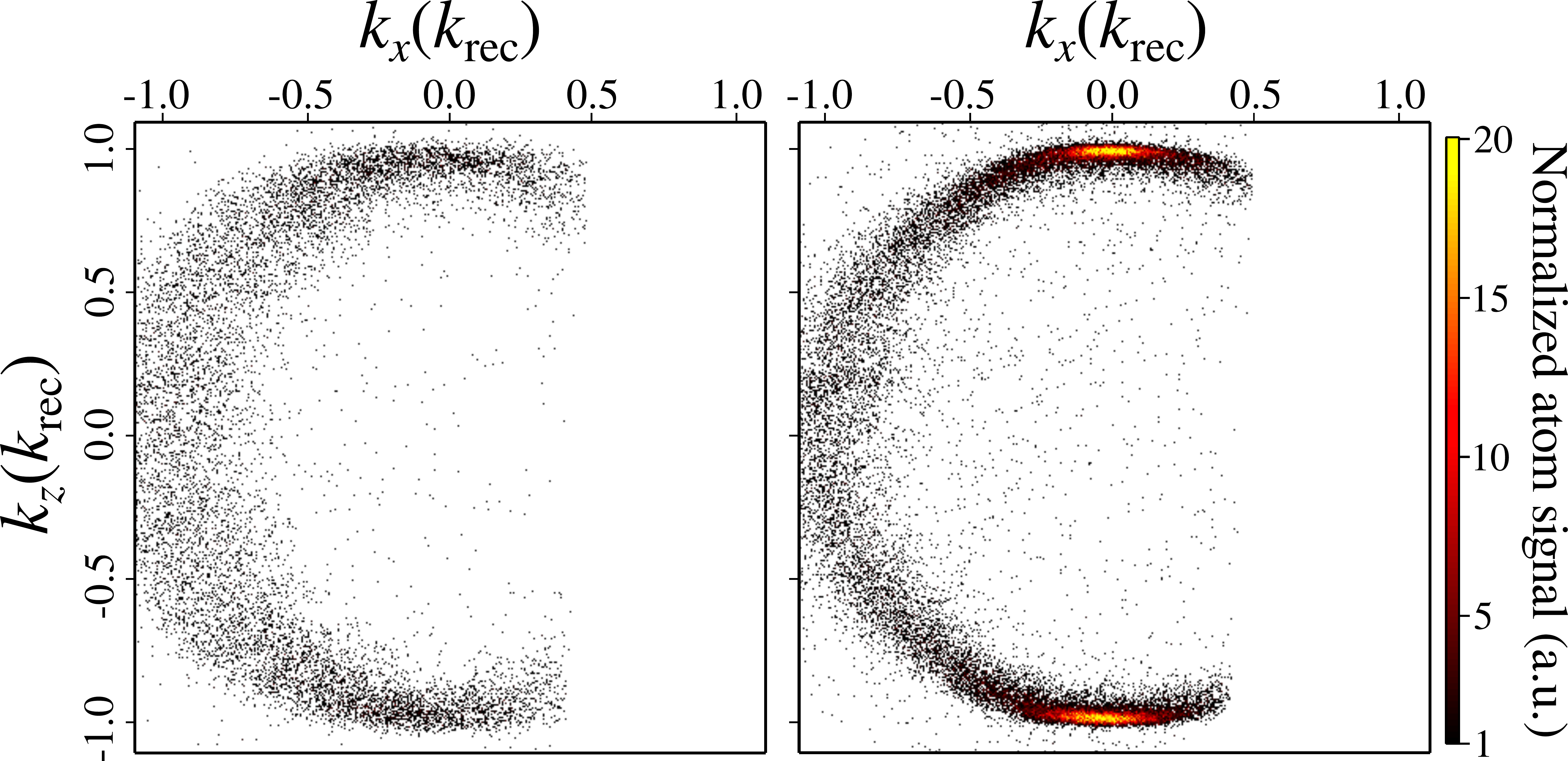}
\caption{(colour online)
Momentum distribution of scattered atoms in the plane perpendicular to the emission dipole.
Both figures show the atom distribution in the $xz$-plane, integrated between $k_y=\pm 0.1\,\krec$ and summed over 100 shots. 
Left: Excitation laser applied 500\,$\mu$s after the trap has been switched off.
Only the radiation pattern for a $y$-polarized dipole is visible.
Right: Excitation laser applied immediately after the trap has been switched off. Strong superradiance is visible in the vertical, endfire modes.
}
\label{2D_X}
\end{center}
\end{figure}

\paragraph{Distribution of atoms in the $xz$ plane} 
The distribution of scattered atoms in the $yz$ plane showed a vanishing population along the direction of the emission dipole (angles 0 and $\pi$ in the Fig.~3 of the main text).
In the $xz$ plane on the other hand, the angular distribution is, as expected, uniform between the superradiant peaks, see Fig.~\ref{AngularPlot_X}.
The signal is zero on one side of each superradiant peak because the atomic cloud in the $xz$ plane is off center with respect to the detector due to the recoil from the excitation laser and the part of the distribution corresponding to $k_x > 0.4\,\krec$ misses the detector as shown in Fig.~\ref{2D_X}.

\paragraph{Calculation of the correlation functions}
 The quantity actually displayed in Fig.~4 of the main text is not the correlation function as defined in Eq.~1, but the one defined by Eq.~\ref{EqS}:

\begin{eqnarray}
\tilde{g}^{(2)}(\Delta {k_z})=\hspace{-2mm} \int_{\Omega_1} \hspace{-3mm} d\Delta k_x d\Delta k_y \int_{\Omega_V}\hspace{-4mm} d^{3}\bf{k} {\langle\,:\!\hat n({\bf k})\hat n({\bf k}+\Delta {\bf k})\!:\,\rangle
\over
\langle \hat n({\bf k})\rangle \;\langle \hat n({\bf k}+\Delta {\bf k}) \rangle} \; \nonumber \\
\tilde{g}^{(2)}(\Delta {k_y})=\hspace{-2mm} \int_{\Omega_2} \hspace{-3mm} d\Delta k_x d\Delta k_z \int_{\Omega_V} \hspace{-4mm} d^{3}\bf{k} {\langle\,:\!\hat n({\bf k})\hat n({\bf k}+\Delta {\bf k})\!:\,\rangle
\over
\langle \hat n({\bf k})\rangle \;\langle \hat n({\bf k}+\Delta {\bf k}) \rangle} \;
\label{EqS}
\end{eqnarray}

The volume $\Omega_1$ is defined by the boundary conditions $|\Delta k_x|<3\times10^{-2}~\krec$, $|\Delta k_y|<3\times10^{-2}~\krec$ and $\Omega_2$ by $|\Delta k_x|<3\times10^{-2}~\krec$, $|\Delta k_z|<3\times10^{-3}~\krec$.
Integration in momentum space is performed over a specific volume $\Omega_V$ for each of the three cases showed in Fig.~4:
\begin{itemize}
\setlength{\parskip}{0em}
\item superradiant peaks case: $|k_x|<0.5~\krec$, $|k_y|<0.5~\krec$ and $|k_z|>0.95~\krec$;
\item scattered sphere away from the superradiant peaks: $|k_z|<0.92~\krec$ and no constraint in the $xy$ plane;
\item stimulated Raman transfer: $\Omega_V$ is the volume centered on the cloud with a width along $z$ of $0.1~\krec$ and no constraint in the $xy$ plane.
\end{itemize}

\begin{figure}[t]
\begin{center}
\includegraphics[width=\columnwidth]{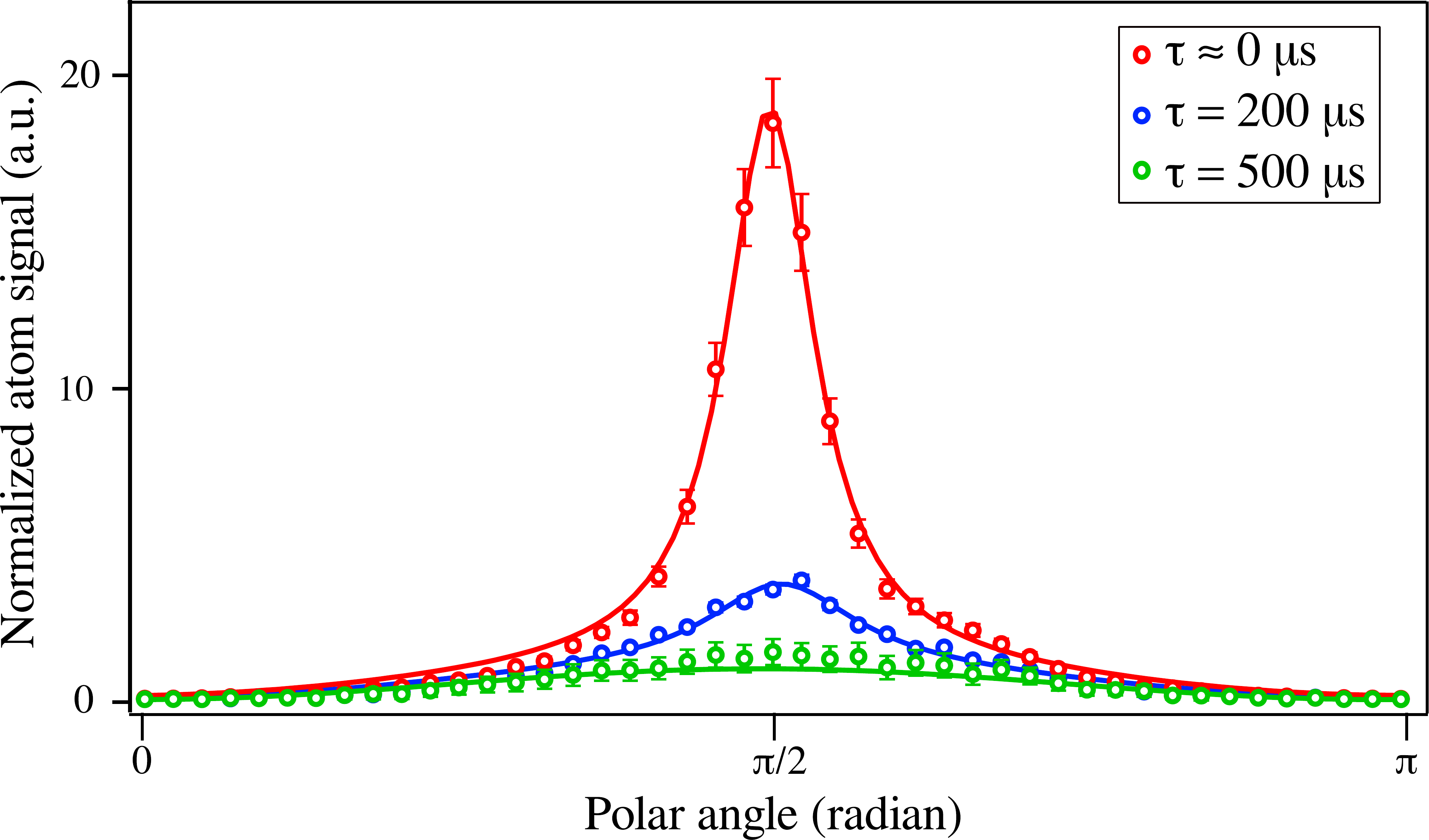}
\caption{(colour online)
Close up of the momentum distribution of scattered atoms around one superradiant peak in the plane of the emission dipole ($yz$ plane). The data shown are the same as those discussed in the main text. Plain lines are fits to the data (see text for details).
}
\label{AngularPlot_Fit}
\end{center}
\end{figure}

\paragraph{Widths of the superradiant peaks}
In order to obtain the widths of the superradiant peak, we first derive the contribution of ordinary spontaneous emission.
Fig.~\ref{AngularPlot_Fit} displays a close up of the superradiant peak around $-\pi/2$ in the $yz$ plane. The data corresponding to a long delay before application of the excitation pulse (green circles, $\tau = 500\,\mu$s) are well described by a pure spontaneous emission profile $\sin^{2}(\theta)$, where $\theta$ is the polar angle in the $yz$ plane (green curve).
Since the contribution of the spontaneous emission should be the same for all delays, we subtract this background from the atomic signal before fitting the distribution with a Lorentzian function. The sum of the background and the fit is also displayed in Fig.~\ref{AngularPlot_Fit} (blue and red curves). The choice of a Lorentzian fitting function is empirical and we expect the exact shape of the superradiant contribution to be more complex \cite{Wasak:2012}.
From this fit we obtain half-widths at half-maximum of $0.14$ and $0.25$\,rad for $\tau = 0$ and $200\,\mu$s, respectively.

Table~\ref{Table_width} summarizes the various widths measured in this experiment. 
The first three lines refer to the widths of the observed atomic distribution in momentum space. 
The "BEC" entry corresponds to the configuration in which the $m=0$ sublevel of the $2^{3}S_{1}$ state was populated by stimulated Raman transfer (see main text).

\begin{table}[h]
\caption{Half-widths at half-maximum for density and correlation function in units of $\krec$. The number in parenthesis denotes the uncertainty on the last digit.}
\label{Table_width}
\begin{center}
\begin{tabular}{|p{5.9cm} r r r|}
   \hline
  Configuration & vertical & \hspace{1mm}& horizontal\\
  \hline
  BEC density  & 0.039(1) & & 0.190(2)\\
 Superradiance density, $\tau = 0$ & 0.014(2) & & 0.14(2)\\
Superradiance density, $\tau = 200~\mu$s & 0.032(2) & & 0.27(2)\\  
Superradiance correlation, $\tau = 0$ & 0.021(2) & & 0.15(1)\\
Scattered sphere correlation, $\tau = 0$ & 0.014(2) & & 0.12(2)\\
\hline
\end{tabular}
\end{center}
\end{table}

\bibliographystyle{apsrev}
\bibliography{superradiance-v6}
\end{document}